\documentclass[aps,pra,showpacs,superscriptaddress,onecolumn]{revtex4}
\usepackage{amsmath}
\usepackage{graphicx}

\begin{document}

\begin{titlepage}
\title
{Dynamics and quantum Zeno effect for a qubit in either
a low- or high-frequency bath beyond the
rotating-wave approximation}
\author{Xiufeng Cao\footnote{Email: xfcao@xmu.edu.cn}}
\affiliation{Advanced Science Institute, The Institute of Physical
and Chemical Research (RIKEN), Wako-shi 351-0198, Japan}
\affiliation{Department of Physics and Institute of Theoretical
Physics and Astrophysics, Xiamen University, Xiamen, 361005,China}
\author{J. Q. You}
\affiliation{Advanced Science Institute, The Institute of Physical
and Chemical Research (RIKEN), Wako-shi 351-0198, Japan}
\affiliation{Department of Physics and Surface Physics Laboratory
(National Key Laboratory), Fudan University, Shanghai 200433, China}
\author{H. Zheng}
\affiliation{Department of Physics, Shanghai Jiao Tong University,
Shanghai 200240, China}
\author{Franco Nori}
\affiliation{Advanced Science Institute, The Institute of Physical
and Chemical Research (RIKEN), Wako-shi 351-0198, Japan}
\affiliation{Physics Department, The University of Michigan, Ann
Arbor, Michigan 48109-1040, USA}
\date{\today }

\begin{abstract}
We use a non-Markovian approach to study the decoherence dynamics of
a qubit (with and without measurement) in either a low- or
high-frequency bath modeling the qubit environment. This approach is
based on a unitary transformation and does not require the
rotating-wave approximation. We show that without measurement, for
low-frequency noise, the bath shifts the qubit energy towards higher
energies (blue shift), while the ordinary high-frequency cutoff
Ohmic bath shifts the qubit energy towards lower energies (red
shift). In order to preserve the coherence of the qubit, we also
investigate the dynamic of qubit with measurement (quantum Zeno
effect) in this two cases: low- and high-frequency baths. For very
frequent projective measurements, the low-frequency bath gives rise
to the quantum anti-Zeno effect on the qubit. The quantum Zeno
effect only occurs in the high-frequency cutoff Ohmic bath, after
considering counter-rotating terms. For a high-frequency
environment, the decay rate should be faster (without measurements)
or slower (with frequent measurements, in the Zeno regime), compared
to the low-frequency bath case. The experimental implementation of
our results here could distinguish the type of bath (either a low-
or high-frequency one) and protect the coherence of the qubit by
modulating the dominant frequency of its environment.
\\

\end{abstract}

\pacs{~03.65.Yz,~03.67.-a,~85.25.-j} \maketitle

\end{titlepage}

\section{Introduction}

There is considerable interest in low-frequency noise (see, e.g., the review
in Ref.~[\onlinecite{add1}] and references therein), because this type of
noise limits the coherence of qubits based on superconducting devices such
as flux or phase qubits \cite{phys-today58-11-42,add6}. Also, the dephasing
of flux qubits is due to low-frequency flux noise with intensity comparable
to the one measured in dc SQUIDs (e.g., Refs.~[%
\onlinecite{add1,prl-97-167001,prl-98-047004}]). Under certain conditions,
noise can enhance the coherence of superconduting flux qubit \cite{prb80}.
There are also several models (e.g., Refs.~[%
\onlinecite{prl-98-267003,prb-76-245306,prl-100-227005}]) for the
microscopic origin of low-frequency flux noise in\ Josephson circuits.
Therefore, the study of low-frequency noise has becomes very important for
superconductor qubits \cite{add1}. Moreover, the quantum Zeno effect has
been proposed as a strategy to protect coherence \cite{A10}, entanglement
\cite{prl-100-090503,A12}, and to control thermodynamic evolutions \cite%
{nature-452-724}. The quantum Zeno effect and anti-Zeno effect have been
widely discussed \cite{Nature-405-546,A11}. So it is an interesting topic to
investigate the quantum Zeno effect of a qubit coupled to a low-frequency
bath.

The description of low-frequency noise \cite{add1} (such as 1/f noise) is
complicated by the presence of long-time correlations in the fluctuating
environment, which prohibit the use of the Markovian approximation. In
addition, the rotating wave approximation (RWA) is also unavailable in an
environment with multiple modes. In the case of a time-depend external field
for a qubit coupled to a thermal bath, Kofman and Kurizki developed a theory
\cite{addKofman} which considers the counter-rotating terms of the fast
modulation field through the negative-frequency part $G(\omega )$ ($\omega
<0 $) in the bath-correlation function spectrum. The dynamics in two
significant models (the spin-boson model with Ohmic bath, and a qubit
coupled to a bath of two level fluctuators) have been calculated within a
rigorous Born approximation and without the Markovian approximation \cite%
{prb-71-035318,prb-79-125317}. Refs.~[\onlinecite{prb-71-035318}] and [%
\onlinecite{prb-79-125317}] describe the structure of the solutions in the
complex plane with branch cuts and poles.

Here we present an analytical approach, based on a unitary transformation.
We use neither the Markovian approximation nor the RWA in order to discuss
the transient dynamics of a qubit coupled to its environment. This method
has already been used \cite{add2} to study the decoherence of the Ohmic
bath, sub-Ohmic bath, and structured bath. In this paper, we calculate the
coherence dynamics of the qubit (with and without measurement) respectively
in two kinds of baths. Besides producing an energy shift, the environment
can change the decay rate of the qubit. To preserve the coherence, we also
investigate the decay rate of the qubit subject to the quantum Zeno effect.
The low-frequency noise uses a Lorentzian-type spectrum with the peak of the
spectrum in the low-energy region and for the high-frequency noise we choose
the ordinary Ohmic bath with Drude cutoff.

Our results show that for \textit{low}-frequency noise, the qubit energy
increases (blue shift) and an anti-Zeno effect takes place. For a \textit{%
high}-frequency cutoff Ohmic bath, the qubit energy decreases (red shift)
and the Zeno effect dominates. For a high-frequency environment, the decay
rate should be faster (without measurements) or slower (with frequent
measurements, in the Zeno regime), compared to the low-frequency bath case.
The experimental implementation of our results here could distinguish the
type of bath (either a low- or high-frequency one) and protect the coherence
of the qubit by modulating the dominant frequency of its environment. The
coherence dynamics without measurement also indicates that the coherence
time of the low-frequency noise is longer than the high-frequency noise,
which demonstrates the powerful temporal memory of the low-frequency bath
and stem from non-Markov process of qubit-bath interaction within the bath
memory time\cite{addnjp}.

\section{Method Beyond the RWA Based on a Unitary Transformation}

We describe a qubit coupled to a boson bath, modeling the environment, by
the Hamiltonian
\begin{equation}
H=-\frac{1}{2}\Delta \sigma _{z}+\frac{1}{2}\sum_{k}g_{k}(a_{k}^{+}+a_{k})%
\sigma _{x}+\sum_{k}\omega _{k}a_{k}^{+}a_{k}.
\end{equation}%
The boson bath and the spin-$\frac{1}{2}$ (fluctuators) bath will have the
same dissipative effect on the qubit at zero temperature, $T=0,$ if both
baths have the same correlation function \cite{U-weiss}. However, for finite
temperatures, the spin bath has a smaller effect on the qubit because of the
likely saturation of the populations in the spin bath. Here, for simplicity,
we only consider the case of zero temperature. That is, our results are also
applicable to a fluctuator-bath at zero temperature. Our approach is based
on a unitary transformation and can be used for different types of
environmental baths. Below we give detailed derivations for the
low-frequency noise case.

The spectral density $J(\omega )$ of the environment considered here is
given by
\begin{equation}
J(\omega )=\sum_{k}g_{k}^{2}\delta (\omega -\omega _{k})=\frac{2\alpha
\omega }{\omega ^{2}+\lambda ^{2}},  \label{E2}
\end{equation}%
where $\lambda $ is an energy lower than the qubit two-energy spacing $%
\Delta $ and $\alpha $ describes the coupling strength between the qubit and
the environment. Choosing $\Delta $ as the energy unit, $\alpha /\Delta ^{2}$
is a dimensionless coupling strength. When $\omega \geq \lambda $, $J(\omega
)\sim 1/\omega $, corresponding to a $1/f$ noise.

We apply a canonical transformation to the Hamiltonian $H$: $H^{\prime
}=\exp \left[ S\right] H\exp \left[ -S\right] .$ Further explanations on the
validity of the transformation can be found in the Appendix. Note that $%
\sum_{k}g_{k}^{2}\xi _{k}{}^{2}/(2\omega _{k}^{2})>0$ in Eq.~(\ref{E9}), so $%
0<\exp \left[ -\sum_{k}g_{k}^{2}\xi _{k}{}^{2}/\left( 2\omega
_{k}^{2}\right) \right] <1.$ Then the solution of $\eta $ will be in the
region from $0$ to $1$. Actually, the existence and uniqueness of the
solution of $\eta $ in Eq.~(\ref{E9}) can be used as a criterion for the
validity of our method. The parameter $\eta $ can be regarded as a
renormalization factor of the energy spacing $\Delta $ and is calculated as
\begin{equation}
\eta =\exp \left\{ \alpha \frac{\pi \lambda \eta \;\Delta +\left[ -\lambda
^{2}-\eta ^{2}\Delta ^{2}+(\lambda ^{2}-\ \eta ^{2}\Delta ^{2})\log
\left\vert \frac{\lambda }{\eta \;\Delta }\right\vert \right] }{(\lambda
^{2}+\eta ^{2})^{2}}\right\} .
\end{equation}%
Obviously, $\eta $ is determined self-consistently by the above equation.

Thus, the effective transformed Hamiltonian can be derived as
\begin{equation}
H^{^{\prime }}\approx -\frac{1}{2}\eta \;\Delta \sigma _{z}+\sum_{k}\omega
_{k}a_{k}^{+}a_{k}+\sum_{k}V_{k}(a_{k}^{+}\sigma _{-}+a_{k}\sigma _{+}).
\label{E11}
\end{equation}%
where%
\begin{equation}
V_{k}=\eta \;\Delta \frac{g_{k}\xi _{k}}{\omega _{k}}.
\end{equation}%
Comparing $H^{\prime }$ in Eq.~(\ref{E11}) with the ordinary Hamiltonian in
the rotating-wave approximation (RWA):
\begin{equation}
H_{\mathrm{RWA}}=-\frac{1}{2}\Delta \sigma _{z}+\sum_{k}\omega
_{k}a_{k}^{+}a_{k}+\sum_{k}\frac{g_{k}}{2}(a_{k}^{+}\sigma _{-}+a_{k}\sigma
_{+}),  \label{E13}
\end{equation}%
one can see that the unitary transformation plays the role of renormalizing
two parameters in the Hamiltonian, i.e., the energy spacing $\Delta $ is
renormalized:
\begin{equation}
\Delta \longrightarrow \eta \;\Delta ;
\end{equation}%
and the coupling strength $g_{k}/2$ between the qubit and the bath is
renormalized:
\begin{equation}
\frac{g_{k}}{2}\longrightarrow \left( \frac{2\eta \;\Delta }{\omega
_{k}+\eta \;\Delta }\right) \frac{g_{k}}{2}.
\end{equation}%
Below we study the decoherence dynamics of the qubit and the quantum Zeno
effect using the transformed Hamiltonian $H^{\prime }$.

\subsection{Non-measurement decoherence dynamics}

We diagonalize the transformed Hamiltonian $H^{\prime }$ in the ground state
$\left\vert g\right\rangle =\left\vert \uparrow \right\rangle \left\vert
0_{k}\right\rangle $ and lowest excited states, $\left\vert \downarrow
\right\rangle \left\vert 0_{k}\right\rangle $ and $\left\vert \uparrow
\right\rangle \left\vert 1_{k}\right\rangle ,$ as
\begin{equation}
H^{^{\prime }}=-\frac{1}{2}\eta \;\Delta \left\vert g\right\rangle
\left\langle g\right\vert +\sum_{E}E\left\vert E\right\rangle \left\langle
E\right\vert ,
\end{equation}%
where $\left\vert \uparrow \right\rangle $ and $\left\vert \downarrow
\right\rangle $ are the eigenstates of $\sigma _{z}$, i.e., $\sigma
_{z}\left\vert \uparrow \right\rangle =\left\vert \uparrow \right\rangle ,$ $%
\sigma _{z}\left\vert \downarrow \right\rangle =-\left\vert \downarrow
\right\rangle $ and $\left\vert n_{k}\right\rangle $ denotes the state with $%
n$ bath excitations for mode $k$. The state $\left\vert E\right\rangle $ is
\begin{equation}
\left\vert E\right\rangle =x(E)\left\vert \downarrow \right\rangle
\left\vert 0_{k}\right\rangle +\sum_{k}y_{k}(E)\left\vert \uparrow
\right\rangle \left\vert 1_{k}\right\rangle ,
\end{equation}%
with $x(E)=\left[ 1+\sum_{k}\frac{V_{k}^{2}}{(E+\eta \Delta /2-\omega
_{k})^{2}}\right] ^{-1/2}$ and $y_{k}(E)=\frac{V_{k}}{E+\eta \Delta
/2-\omega _{k}}x(E).$

Here we calculate the dynamical quantity $\left\langle \sigma
_{x}(t)\right\rangle ,$ which\ is the analog of the population inversion $%
\left\langle \sigma _{z}(t)\right\rangle $ in the spin-boson model \cite%
{add3}. Since the coupling to the environment will be \textquotedblleft
always present\textquotedblright\ in essentially all physically relevant
situations, a natural ground state is given by the dressed state of the
two-level qubit and bath. Therefore, the ground state of $H$ is $\exp
[-S]\left\vert \uparrow \right\rangle \left\vert 0_{k}\right\rangle ,$ under
counter-rotating terms, and the corresponding ground-state energy is $-\eta
\;\Delta /2,$ then the ground state of $H^{^{\prime }}$ becomes $\left\vert
\uparrow \right\rangle \left\vert 0_{k}\right\rangle $ with the identical
ground-state energy $-\eta \;\Delta /2$. We now prepare the initial state,
which is also a dressed state of the qubit and bath, from the ground state as%
\begin{equation}
\left\vert \psi (0)\right\rangle =\frac{\left( 1+\sigma _{x}\right) }{\sqrt{2%
}}\exp [-S]\left\vert \uparrow \right\rangle \left\vert 0_{k}\right\rangle .
\end{equation}%
Then the initial state in the \textit{transformed Hamiltonian} becomes $%
\left\vert \psi (0)\right\rangle ^{^{\prime }}=(\left\vert \uparrow
\right\rangle +\left\vert \downarrow \right\rangle )\left\vert
0_{k}\right\rangle /\sqrt{2}$, which is the eigenstate of $\sigma _{x}$.
Starting from this initial state, we obtain \cite{EPJB-2004-H-Zheng}%
\begin{eqnarray}
\left\langle \sigma _{x}(t)\right\rangle &=&\mathrm{Tr}_{B}\left\langle \psi
(t)\left\vert \sigma _{x}\right\vert \psi (t)\right\rangle  \notag \\
&=&\mathrm{Tr}_{B}\left\langle \psi (0)\left\vert \exp [iHt]\sigma _{x}\exp
[-iHt]\right\vert \psi (0)\right\rangle  \notag \\
&=&\frac{1}{2}\sum_{E}x(E)^{2}\exp \left[ -i\left( E+\frac{\eta \;\Delta }{2}%
\right) t\right] +\frac{1}{2}\sum_{E}x(E)^{2}\exp \left[ i\left( E+\frac{%
\eta \;\Delta }{2}\right) t\right]  \notag \\
&=&\frac{1}{4\pi i}\int_{\infty }^{-\infty }dE^{^{\prime }}\exp \left[
-iE^{^{\prime }}t\right] \left( E^{^{\prime }}-\eta \;\Delta -\sum_{k}\frac{%
V_{k}^{2}}{E^{^{\prime }}+i0^{+}-\omega _{k}}\right) ^{-1}  \notag \\
&&+\frac{1}{4\pi i}\int_{-\infty }^{\infty }dE^{^{\prime }}\exp \left[
iE^{^{\prime }}t\right] \left( E^{^{\prime }}-\eta \;\Delta -\sum_{k}\frac{%
V_{k}^{2}}{E^{^{\prime }}-i0^{+}-\omega _{k}}\right) ^{-1}  \notag \\
&=&\mathrm{Re}\left[ \frac{1}{2\pi i}\int_{\infty }^{-\infty }\frac{\exp %
\left[ -iE^{^{\prime }}t\right] }{E^{^{\prime }}-\eta \;\Delta -\sum_{k}%
\frac{V_{k}^{2}}{E^{^{\prime }}+i0^{+}-\omega _{k}}}dE^{^{\prime }}\right] .
\end{eqnarray}%
Here we denote the real and imaginary parts of $\sum_{k}V_{k}^{2}/\left(
\omega -\omega _{k}\pm i0^{+}\right) $ as $R(\omega )$ and $\mp \Gamma
(\omega )$, respectively. It follows that
\begin{eqnarray}
R(\omega ) &=&\wp \sum_{k}\frac{V_{k}^{2}}{\omega -\omega _{k}}=(\eta
\;\Delta )^{2}\wp \int\limits_{0}^{\infty }d\omega ^{^{\prime }}\frac{%
J(\omega ^{^{\prime }})}{(\omega -\omega ^{^{\prime }})(\omega ^{^{\prime
}}+\eta \Delta )^{2}}  \notag \\
&=&2\alpha \Delta ^{2}\eta ^{2}\frac{\omega \log \left\vert \omega
\right\vert }{(\eta \;\Delta +\omega )^{2}(\lambda ^{2}\ +\omega ^{2})}
\notag \\
&&+2\alpha \Delta ^{2}\eta ^{2}\frac{\pi \lambda \left[ \lambda ^{2}+\eta
\;\Delta \ (-\eta \;\Delta +2\omega )\right] -2\left[ \lambda ^{2}(2\eta
\;\Delta -\omega )+\eta ^{2}\Delta ^{2}\ \omega \right] \log \left\vert
\lambda \right\vert }{2(\lambda ^{2}+\eta ^{2}\Delta ^{2})^{2}(\lambda
^{2}+\ \omega ^{2})}  \notag \\
&&+2\alpha \Delta ^{2}\eta ^{2}\frac{-(\eta \;\Delta +\omega )\left( \lambda
^{2}+\Delta ^{2}\eta ^{2}\right) +\left[ 2\eta ^{3}\Delta ^{3}+(-\lambda
^{2}+\eta ^{2}\Delta ^{2})\ \omega \right] \log \left\vert \eta \;\Delta
\right\vert }{(\lambda ^{2}+\eta ^{2}\Delta ^{2})^{2}(\ \eta \;\Delta
+\omega )^{2}},
\end{eqnarray}%
and%
\begin{equation}
\Gamma (\omega )\;=\;\pi \sum_{k}V_{k}^{2}\delta (\omega -\omega
_{k})\;=\;\pi (\eta \;\Delta )^{2}\frac{J(\omega )}{(\omega +\eta \;\Delta
)^{2}},
\end{equation}%
where $\wp $ stands for the Cauchy principal value and $J(\omega )$ is the
spectral density. Then, we have
\begin{equation}
\left\langle \sigma _{x}(t)\right\rangle =\frac{1}{\pi }\int_{0}^{\infty }%
\frac{\Gamma (\omega )\cos \omega t}{(\omega -\eta \;\Delta -R(\omega
))^{2}+\Gamma (\omega )^{2}}d\omega .  \label{E26}
\end{equation}%
The integration in Eq.~(\ref{E26}) can be calculated numerically or
approximately using residual theory.

\subsection{Measurement dynamics: quantum Zeno effect}

It is known that the quantum Zeno effect can effectively slow down the
quantum decay rate of a quantum system. We study this effect using an
approach that goes beyond the Markovian approximation and RWA. Here we
consider the low-frequency bath as in Sec.~II.A and derive the effective
decay rate. The Hamiltonian $H^{^{\prime }}$ is given in Eq.~(11),
\begin{equation}
H^{^{\prime }}\approx -\frac{1}{2}\eta \;\Delta \sigma _{z}+\sum_{k}\omega
_{k}a_{k}^{+}a_{k}+\sum_{k}\eta \;\Delta \frac{g_{k}\xi _{k}}{\omega _{k}}%
(a_{k}^{+}\sigma _{-}+a_{k}\sigma _{+}).
\end{equation}%
In this paper, the effective decay rate is defined in the same form as in
Ref.~[\onlinecite{Nature-405-546}].

Write the wave function in the transformed Hamiltonian as
\begin{equation}
\left\vert \Phi (t)\right\rangle ^{^{\prime }}=\chi (t)\left\vert \downarrow
\right\rangle \left\vert 0_{k}\right\rangle +\sum_{k}\beta _{k}(t)\left\vert
\uparrow \right\rangle \left\vert 1_{k}\right\rangle ,
\end{equation}%
with probability in the excited state at initial time $\left\vert \chi
(0)\right\vert ^{2}$. Substituting $\left\vert \Phi (t)\right\rangle
^{^{\prime }}$ into the Schr\"{o}dinger equation with $H^{^{\prime }}$, we
have
\begin{equation}
i\frac{d\chi (t)}{dt}=\frac{\eta \;\Delta }{2}\chi (t)+\sum_{k}V_{k}\;\beta
_{k}(t),  \label{E29}
\end{equation}%
\begin{equation}
i\frac{d\beta _{k}(t)}{dt}=\left( \omega _{k}-\frac{\eta \;\Delta }{2}%
\right) \beta _{k}(t)+\sum_{k}V_{k}\;\chi (t).  \label{E30}
\end{equation}%
When the transformations
\begin{equation}
\chi (t)\!\!=\!\!\widetilde{\chi }(t)\exp \left[ -i\frac{\eta \;\Delta }{2}t%
\right] ,
\end{equation}%
\begin{equation}
\beta _{k}(t)\!\!=\!\!\widetilde{\beta }_{k}(t)\exp \left[ -i\left( \omega
_{k}-\frac{\eta \;\Delta }{2}\right) t\right]
\end{equation}%
are applied, Eqs.~(\ref{E29}) and (\ref{E30}) can be written as
\begin{equation}
\frac{d\widetilde{\chi }(t)}{dt}=-i\sum_{k}V_{k}\widetilde{\beta }%
_{k}(t)\exp \left[ -i(\omega _{k}-\eta \;\Delta )t\right] ,  \label{E33}
\end{equation}%
\begin{equation}
\frac{d\widetilde{\beta }_{k}(t)}{dt}=-iV_{k}\widetilde{\chi }(t)\exp \left[
i(\omega _{k}-\eta \;\Delta )t\right] .  \label{E34}
\end{equation}%
Integrating Eq.~(\ref{E34}) and then substituting it into Eq.~(\ref{E33}),
we obtain
\begin{equation}
\frac{d\widetilde{\chi }(t)}{dt}=-\sum_{k}V_{k}^{2}\int\limits_{0}^{t}%
\widetilde{\chi }(t^{^{\prime }})\exp [-i(\omega _{k}-\eta \;\Delta
)(t-t^{^{\prime }})]\;dt^{^{\prime }}.  \label{E35}
\end{equation}%
This integro-differential equation Eq.~(\ref{E35}) is exactly soluble by a
Laplace transformation.

As for the present study of the quantum Zeno effect, i.e., using frequent
measurements, it suffices to obtain the short-time behavior and the equation
can be solved iteratively. With the initial excited-state probability
amplitude $\chi (0)$, in the first iteration, Eq.~(\ref{E35}) is solved as
\begin{equation}
\frac{\widetilde{\chi }(t)}{\chi (0)}\simeq
1-\int\limits_{0}^{t}(t-t^{^{\prime }})\sum_{k}V_{k}^{2}\exp [-i(\omega
_{k}-\eta \;\Delta )t^{^{\prime }}]\;dt^{^{\prime }}.
\end{equation}%
We can approximately write $\widetilde{\chi }(t)$ using an exponential form:
\begin{eqnarray}
\frac{\widetilde{\chi }(t)}{\chi (0)} &=&\exp \left[ -\int%
\limits_{0}^{t}(t-t^{^{\prime }})\sum_{k}V_{k}^{2}\exp \left[ -i(\omega
_{k}-\eta \;\Delta )t^{^{\prime }}\right] dt^{^{\prime }}\right]  \\
&=&\exp \left\{ -t\left[ -\frac{1}{t}\sum_{k}V_{k}^{2}\frac{\exp \left[
-i\left( \omega _{k}-\eta \;\Delta \right) t\right] -1+i\left( \omega
_{k}-\eta \;\Delta \right) t}{\left( \omega _{k}-\eta \;\Delta \right) ^{2}}%
\right] \right\}  \\
&=&\exp \left\{ -t\left[ \sum_{k}V_{k}^{2}\left( \frac{2\sin \left( \frac{%
\omega _{k}-\eta \;\Delta }{2}t\right) ^{2}}{t\left( \omega _{k}-\eta
\;\Delta \right) ^{2}}-i\frac{\left( \omega _{k}-\eta \;\Delta \right)
t-\sin \left[ \left( \omega _{k}-\eta \;\Delta \right) t\right] }{t\left(
\omega _{k}-\eta \;\Delta \right) ^{2}}\right) \right] \right\} .
\end{eqnarray}%
The instantaneous ideal projections are assumed to be performed at intervals
$\tau $. If single measurement, the probability amplitude is $\widetilde{%
\chi }(t=\tau ).$ For a sufficiently large frequency of measurements, the
survival population in the excited state is%
\begin{equation}
\rho _{\mathrm{ee}}(t=n\tau )=\left\vert \widetilde{\chi }(t=n\tau
)\right\vert ^{2}=\left\vert \chi (0)\right\vert ^{2}\exp [-\gamma (\tau )t],
\label{E40}
\end{equation}%
where the subscript \textquotedblleft $\mathrm{ee}$\textquotedblright\
refers to the initial and final excited state. And $\gamma (\tau )$, with
projection intervals $\tau ,$ is obtained as
\begin{eqnarray}
\gamma (\tau ) &=&2\pi \int_{0}^{\infty }d\omega \sum_{k}\left( \frac{g_{k}}{%
2}\right) ^{2}\left( \eta \;\Delta \frac{2\xi _{k}}{\omega _{k}}\right) ^{2}%
\frac{2\sin ^{2}(\frac{\eta \;\Delta -\omega }{2}\tau )}{\pi (\eta \Delta
-\omega )^{2}\tau },  \label{E41} \\
&=&2\pi \int_{0}^{\infty }d\omega \frac{J(\omega )}{4}\left[ 1-\frac{\omega
-\eta \;\Delta }{\omega +\eta \;\Delta }\right] ^{2}\frac{2\sin ^{2}(\frac{%
\eta \;\Delta -\omega }{2}\tau )}{\pi (\eta \Delta -\omega )^{2}\tau }.
\label{E42}
\end{eqnarray}%
We now prepare the qubit to the dressed excited state $\exp [-S]\left\vert
\downarrow \right\rangle \left\vert 0_{k}\right\rangle $ $(\sigma
_{z}\left\vert \downarrow \right\rangle =-\left\vert \downarrow
\right\rangle )$ at the initial time $t=0$, which can be achieved by acting
the operator $\sigma _{x}$ on the ground state $\exp [-S]\left\vert \uparrow
\right\rangle \left\vert 0_{k}\right\rangle ,$%
\begin{equation}
\left\vert \Phi (0)\right\rangle =\exp [-S]\left\vert \downarrow
\right\rangle \left\vert 0_{k}\right\rangle =\sigma _{x}\exp [-S]\left\vert
\uparrow \right\rangle \left\vert 0_{k}\right\rangle .
\end{equation}%
In this case, the initial state in the transformed Hamiltonian is $%
\left\vert \downarrow \right\rangle \left\vert 0_{k}\right\rangle $ and $%
\chi (0)=1.$

Note that in Eq.~(\ref{E42}), the renormalization factor $\eta $ of the
characteristic energy $\Delta $ appears in the decay rate $\gamma (\tau )$.
This is different from the formulas for $\gamma (\tau )$ in Refs.~[%
\onlinecite{PRL-101-200404}] and [\onlinecite{pRA-80-023801}]. In the case
of spontaneous emission, the coupling strength between the electromagnetic
field and atom is the fine structure constant $1/137$, so it belongs to the
weak-coupling case and $\eta $ then becomes extremely close to $1$.
Therefore, besides spontaneous emission, this result for the quantum Zeno
effect can apply to other cases of strong coupling between the qubit and the
bath. \ \ \

Under the RWA, the Hamiltonian becomes $H_{\mathrm{RWA}}$ as in Eq.~(\ref%
{E13}). We prepare the initial excited state of $H$ through the operator $%
\sigma _{x}$ acting on the ground state under RWA $\left\vert \uparrow
\right\rangle \left\vert 0_{k}\right\rangle ,$ $\sigma _{x}\left\vert
\uparrow \right\rangle \left\vert 0_{k}\right\rangle =\left\vert \downarrow
\right\rangle \left\vert 0_{k}\right\rangle $. Then, following the
derivation in Refs.~[\onlinecite{Nature-405-546}], the decay rate is reduced
to
\begin{eqnarray}
\gamma _{\mathrm{RWA}}(\tau ) &=&2\pi \int_{0}^{\infty }d\omega
\sum_{k}\left( \frac{g_{k}}{2}\right) ^{2}\frac{2\sin ^{2}(\frac{\Delta
-\omega }{2}\tau )}{\pi (\Delta -\omega )^{2}\tau }. \\
&=&2\pi \int_{0}^{\infty }d\omega \frac{J(\omega )}{4}\frac{2\sin ^{2}(\frac{%
\Delta -\omega }{2}\tau )}{\pi (\Delta -\omega )^{2}\tau }.
\end{eqnarray}

To compare with the high-frequency Ohmic bath, we choose the ordinary Ohmic
bath with Drude cutoff:%
\begin{equation}
J^{\mathrm{oh}}(\omega )=\sum_{k}g_{k}^{2}\delta (\omega -\omega _{k})=\frac{%
2\alpha ^{\mathrm{oh}}\omega }{\left( \omega /\omega _{c}\right) ^{2}+1}.
\label{E46}
\end{equation}%
This is a realistic assumption for, e.g., an electromagnetic noise. In Eq.~(%
\ref{E46}), $\alpha ^{\mathrm{oh}}$ is the coupling strength between the
qubit and the Ohmic bath. The cutoff frequency $\omega _{c}$ in the spectral
density $J^{\mathrm{oh}}(\omega )$ is typically assumed to be the largest
frequency in the problem. As for the low-frequency noise, we use $J(\omega
)=2\alpha \omega /(\omega ^{2}+\lambda ^{2})$, which is the same as in Eq.~(%
\ref{E2}). The difference between these two baths is that $\lambda $
corresponds to an energy lower than the qubit energy. In the next section,
we will show our numerical results for these two baths.

\section{Results and Discussions}

To show the effects of either a low- or a high-frequency noise on the qubit
states respectively, we study the dynamical quantities $\left\langle \sigma
_{x}(t)\right\rangle $ and the quantum Zeno decay rate $\gamma (\tau )$. The
energy shift and the quantum Zeno decay rate exhibit evidently different
features for the low- and high-frequency noises. Thus, these two quantities
(energy shift and decay rate) can be used as criteria to \textit{distinguish
the type of noise}. In Fig.~1, we show the spectral densities $J(\omega )$
and $J^{\mathrm{oh}}(\omega )$ of the two baths in the cases of both weak
dissipation $\alpha /\Delta ^{2}=0.01$ ($\alpha ^{\mathrm{oh}}=0.01$) and
strong dissipation $\alpha /\Delta ^{2}=0.1$ ($\alpha ^{\mathrm{oh}}=0.1$).
As examples, we choose $\lambda =0.09\Delta $ and $\lambda =0.3\Delta $ in
Fig.~1(a) and 1(b), respectively. Here, each value of $\lambda $ corresponds
to the position of the peak in the low-frequency spectral density. For an
Ohmic bath, the cutoff frequency is fixed at $\omega _{c}=10\Delta $. As
expected, these very different low- and high-frequency spectral densities
should give rise to different decoherence behaviors of the qubit. In Fig.~1,
we also show the characteristic energy of the isolated qubit $\Delta $ (see
the vertical dotted line in Fig.~1).

\begin{figure}[tbp]
\includegraphics[width=9cm,clip]{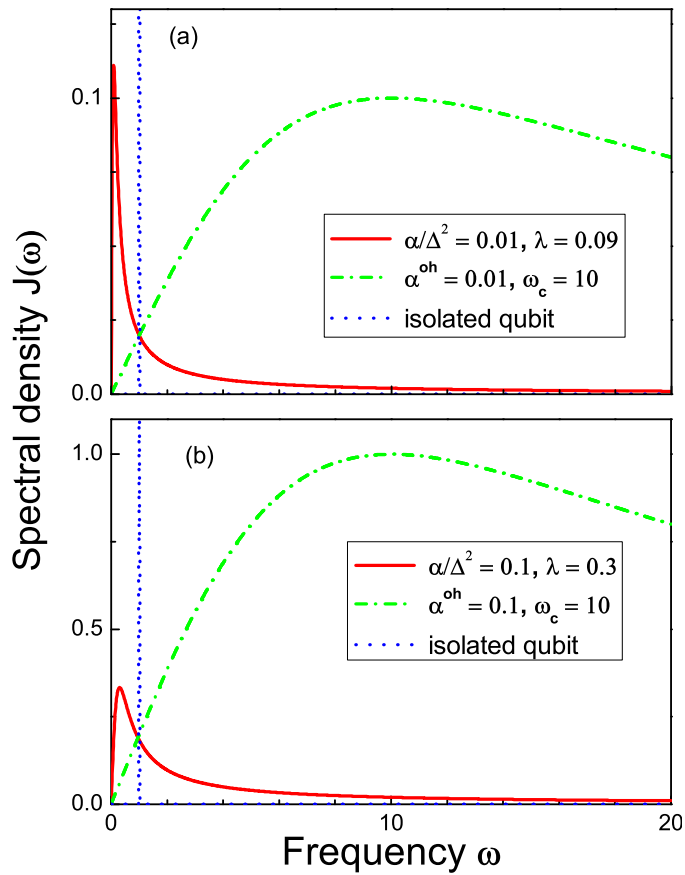}
\caption{(Color online) The spectral density $J(\protect\omega)$ of the low-
and high-frequency baths. (a)~The case of \textit{weak} interaction between
the bath and the qubit, where the parameters of the low-frequency
Lorentzian-like spectrum are $\protect\alpha /\Delta^{2}=0.01$ and $\protect%
\lambda =0.09\Delta $ (red solid curve), while for the high-frequency Ohmic
bath with Drude cutoff the parameters are $\protect\alpha ^{\mathrm{oh}%
}=0.01 $ and $\protect\omega _{c}=10\Delta $ (green dashed-dotted curve).
(b)~The case of \textit{strong} interaction between the bath and the qubit,
where the parameters of the low-frequency bath are $\protect\alpha /\Delta
^{2}=0.1 $ and $\protect\lambda =0.3\Delta$ (red solid curve) and the
parameters of the high-frequency Ohmic bath are $\protect\alpha ^{\mathrm{oh}%
}=0.1$ and $\protect\omega _{c}=10\Delta$ (green dashed-dotted curve). The
characteristic energy of the isolated qubit is indicated by a vertical blue
dotted line. Here, and in the following figures, the energies are shown in
units of $\Delta $.}
\end{figure}

\subsection{Non-measurement decoherence dynamics}

Results from the numerical integration of Eq.~(26) are shown in Fig.~2. They
are qualitatively consistent with the results obtained using residual
theory. This indicates that the branch cuts considered in Refs.~%
\onlinecite{prb-71-035318} and \onlinecite{prb-79-125317} do not affect the
oscillation frequency. The time evolution of $\left\langle \sigma
_{x}(t)\right\rangle $ is given in Fig.~2(a) for the case of weak coupling
between the qubit and the bath, where $\alpha /\Delta ^{2}=0.01$ and $%
\lambda =0.09\Delta $ for the low-frequency noise and $\alpha ^{\mathrm{oh}%
}=0.01$ and $\omega _{c}=10\Delta $ for the Ohmic bath. Figure~2(b) presents
the time evolution of $\left\langle \sigma _{x}(t)\right\rangle $ in the
strong coupling case with the parameters $\alpha /\Delta ^{2}=0.1$ and $%
\lambda =0.3\Delta $ for the low-frequency noise, as well as $\alpha ^{%
\mathrm{oh}}=0.1$ and $\omega _{c}=10\Delta $ for the ohmic bath. As
expected, the quantum oscillations of $\left\langle \sigma
_{x}(t)\right\rangle $ dampen faster in the strong-coupling case. We
approximately evaluate the oscillation frequency or the effective energy of
the qubit, $\omega _{0}-\eta \Delta -R(\omega _{0})=0$, using the residue
theorem. The decay rate can be obtained from $\Gamma (\omega )$. We will now
show the numerical values of $\eta $ in the corresponding cases. In
Fig.~2(a), the renormalized factor $\eta =0.98336$, the oscillation
frequency is $\omega _{0}=1.0225\Delta $ and the decay rate is $\Gamma
(\omega _{0})=0.014654\Delta $ for the low-frequency noise; while $\eta
=0.98447,$ $\omega _{0}=0.97720\Delta ,$ and $\Gamma (\omega
_{0})=0.015318\Delta $ for the Ohmic bath. In Fig.~2(b), the renormalized
factor $\eta =0.91444$, the oscillation frequency is $\omega
_{0}=1.0868\Delta $ and the decay rate is $\Gamma (\omega
_{0})=0.11215\Delta $ for the low-frequency noise, while $\eta =0.84469$, $%
\omega _{0}=0.77221\Delta $, and $\Gamma (\omega _{0})=0.13163\Delta $ for
the Ohmic bath.

\begin{figure}[tbp]
\includegraphics[width=9cm,clip]{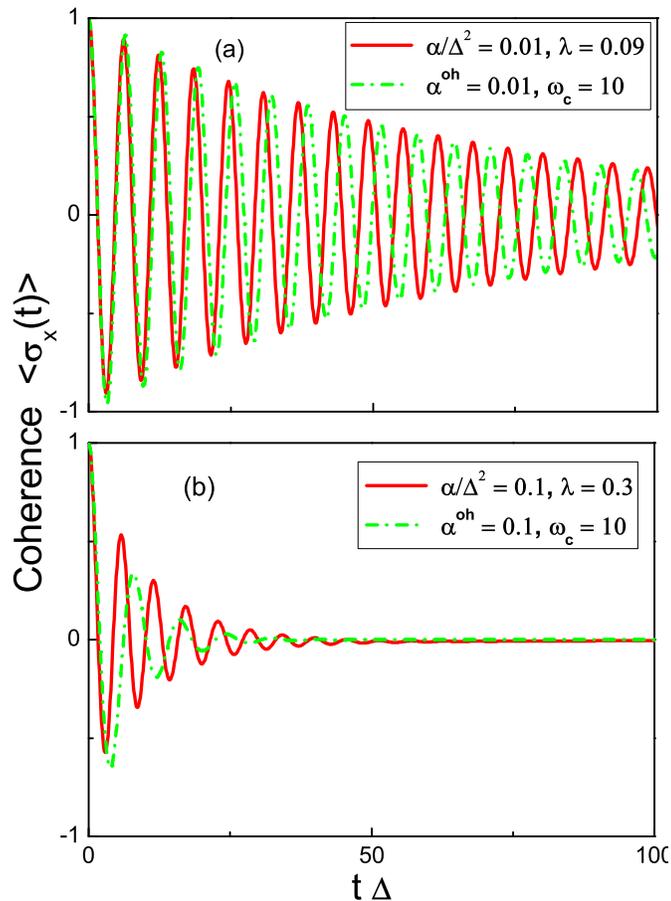}
\caption{(Color online) Time evolution of the coherence $\left\langle
\protect\sigma _{x}(t)\right\rangle $ versus the time $t$ multiplied by the
qubit energy spacing $\Delta$. (a) The case of \textit{weak} interaction
between the bath and the qubit, where the parameters of the low-frequency
Lorentzian-type spectrum are $\protect\alpha /\Delta ^{2}=0.01, ${\ }$%
\protect\lambda =0.09\Delta ${\ (red solid curve); while for the
high-frequency Ohmic bath with Drude cutoff the parameters are }$\protect%
\alpha ^{\mathrm{oh}}=0.01,$ $\protect\omega _{c}=10\Delta $ (green
dashed-dotted curve){. (b) }The case of \textit{strong} interaction between
the bath and the qubit, where the parameters of the low-frequency bath are $%
\protect\alpha /\Delta ^{2}=0.1,${\ }$\protect\lambda =0.3${\ (red solid
line)}, and for the high-frequency Ohmic bath are{\ }$\protect\alpha ^{%
\mathrm{oh}}=0.1,$ $\protect\omega _{c}=10\Delta $ (green dashed-dotted
line). These results show that the decay rate for the low-frequency bath is
\textit{shorter} than for the high-frequency Ohmic bath. This means that the
coherence time of the qubit in the low-frequency bath is \textit{longer}
than in the high-frequency noise case, demonstrating the powerful temporal
memory of the low-frequency bath. Also, our results reflect the structure of
the solution with branch cuts~\protect\cite{prb-71-035318}. The oscillation
frequency for the low-frequency noise is $\protect\omega _{0}>\Delta $, in
spite of the strength of the interaction. This can be referred to as a blue
shift. However, in an Ohmic bath, the oscillation frequency is $\protect%
\omega _{0}<\Delta $, corresponding to a red shift. The shifting direction
of the energy is independent of the interaction strength and only determined
by the spectral properties. Thus, it can be used as a \textit{criterion for
distinguishing the low- and high-frequency noises}.}
\end{figure}

The energy spectral densities in Fig.~1 and the results in Fig.~2 indicate
\textit{two opposite shifts} of the characteristic energy $\Delta $ for the
two kinds of baths considered here. These opposite energy shifts are
equivalent to \textit{energy repulsion}. The energy shift is determined by
the interaction term. In the transformed Hamiltonian, the interaction is $%
H_{1}^{\prime }$ in Eq.~(\ref{E7}). Also, dipolar interactions, such as $%
H_{I}^{JC}=g(a^{+}\sigma _{-}+a\sigma _{+})$ in the Jaynes-Cummings model,
decrease the qubit's ground-state energy and increase its excited-state
energy. Thus, now we ask the following questions: how a qubit is affected by
either a multimode bath or a single-mode cavity? How a qubit is influenced
by these two kinds of multimode baths: low-frequency and high-frequency ones?

For a low-frequency bath, the energy peak of the bath is located between the
ground-state energy and the excited-state energy, that is to say the main
part of the spectrum is in the region $\omega _{k}<\Delta $. Then (as seen
in Fig.~1) the interaction of the bath with the two qubit states is
\textquotedblleft opposite\textquotedblright , i.e. the ground-state energy
becomes \textit{lower} and the excited-state energy becomes \textit{higher}.
So the energy spacing for the case of a low-frequency bath exhibits a
\textit{blue shift}. This result is similar to the single-mode
Jaynes-Cummings model. The energy difference of the two-state qubit is
\textit{increased} by the low-frequency bath.

For a high-frequency cutoff Ohmic bath, the energy peak of the bath is
located above the excited-state energy. So the main part of the spectrum is
in the region $\omega _{k}>\Delta $. The effect of the bath on the qubit
mainly comes from the frequencies higher than the excited-state energy\ of
the qubit. Then (as seen in Fig.~1) the bath repels both the excited-state
and the ground-state energies to lower energies.\ But the effect of the
high-frequency bath on the excited state is much larger than on the ground
state. As a result, on the whole, the qubit energy difference in a
high-frequency bath is \textit{red shifted}. Thus, the effective energy
difference of the qubit is \textit{reduced} by the high-frequency bath.

For example, if the initial state of the qubit is an excited state, in the
interaction picture, the main part of the coupling is
\begin{eqnarray}
&&\!\sum_{k}g_{k}a_{k}^{+}\exp (i\omega _{k}t)\sigma _{-}\exp (-i\Delta t)
\notag \\
=\! &&\!\sum_{k}g_{k}a_{k}^{+}\sigma _{-}\left\{ \cos [(\omega _{k}-\Delta
)t]+i\sin [(\omega _{k}-\Delta )t]\right\} ,
\end{eqnarray}%
where the real part contributes to the decay rate and the imaginary part
results in the energy shift. For the low-frequency bath, the main part of
the spectrum is in the region $\omega _{k}<\Delta $. Thus, the term for the
energy shift is
\begin{equation}
\sin [(\omega _{k}-\Delta )t]<0.
\end{equation}%
However, for a high-frequency cutoff Ohmic bath, the main part of the
spectrum is in the region $\omega _{k}>\Delta $, where the term for energy
shift becomes
\begin{equation}
\sin [(\omega _{k}-\Delta )t]>0.
\end{equation}%
These results show that the energy shift for the two kinds of baths moves in
opposite directions. Note that there is a minus in the interaction term in
the expressions for the dynamical quantities such as Eq.~(21) and Eq.~(33).
The above observations help us understand \textit{why} the energy levels
repel. The contribution by the real part of the interaction on the decay
rate will be discussed below, when studying the quantum Zeno effect.

\subsection{Measurement dynamics: quantum Zeno effect}

The quantum Zeno effect can be a useful tool to preserve the state coherence
of a quantum system, with the help of repeated projective measurements.
Below we investigate the quantum Zeno effect in the qubit system and \textit{%
propose another criterion for distinguishing low- and high-frequency noises}%
. In general, without using the RWA, the effective decay rate can be
obtained as
\begin{equation}
\gamma (\tau )=2\pi \int_{0}^{\infty }d\omega J(\omega )\left( 1-\frac{%
\omega -\eta \;\Delta }{\omega +\eta \;\Delta }\right) ^{2}\frac{2\sin ^{2}(%
\frac{\eta \;\Delta -\omega }{2}\tau )}{\pi (\eta \;\Delta -\omega )^{2}\tau
}.
\end{equation}%
This expression includes three terms, i.e., the spectral density $J(\omega )$
of the bath, the projection time modulating function
\begin{equation}
F(\omega ,\tau )=\frac{2\sin ^{2}(\frac{\eta \;\Delta -\omega }{2}\tau )}{%
\pi (\eta \;\Delta -\omega )^{2}\tau },
\end{equation}%
and the interaction contribution function of both the rotating and
counter-rotating terms
\begin{equation}
f(\omega )=\left( 1-\frac{\omega -\eta \;\Delta }{\omega +\eta \;\Delta }%
\right) ^{2}.
\end{equation}%
The counter-rotating term contributing to $f(\omega )$ is $(\omega -\eta
\;\Delta )/\left( \omega +\eta \;\Delta \right) $. If the RWA is applied, $%
f(\omega )=1$. In Fig. 3(a), the decay rate $\gamma (\tau )/\gamma _{0}$ for
the low-frequency bath is plotted in the weak-coupling case with $\alpha
/\Delta ^{2}=0.01$. Here $\gamma _{0}$ is the decay rate for $\left\vert
e\right\rangle \rightarrow \left\vert g\right\rangle $ in the long-time
limit with the RWA,
\begin{equation}
\gamma _{0}=\gamma (\tau \rightarrow \infty )=2\pi J(\Delta )/4.
\end{equation}%
From the energy spectrum in Fig.~1(a), we can see that $\gamma _{0}$ is
proportional to the spectrum density of the bath, with the magnitude
corresponding to the crossing of the energy $\Delta $ of the isolated qubit
and the bath spectrum. For comparison, we also plot $\gamma _{\mathrm{RWA}%
}(\tau )/\gamma _{0}$ as a dashed-dotted curve, with
\begin{equation}
\gamma _{\mathrm{RWA}}(\tau )=2\pi \int_{0}^{\infty }d\omega \frac{J(\omega )%
}{4}\frac{2\sin ^{2}(\frac{\Delta -\omega }{2}\tau )}{\pi (\Delta -\omega
)^{2}\tau }.
\end{equation}%
As we know, $\gamma (\tau )/\gamma _{0}<1$ means that repeated measurements
slow down the decay rate $\gamma (\tau )<\gamma _{0},$ which is the quantum
Zeno effect. On the contrary, $\gamma (\tau )/\gamma _{0}>1$ means an
anti-Zeno effect. The curves in Fig.~3(b) show results for the Ohmic bath
with $\alpha ^{\mathrm{oh}}=0.01$. In Fig.~4, we show the decay rate in the
strong-coupling case for the low-frequency bath with $\alpha /\Delta
^{2}=0.1 $ and for the Ohmic bath with $\alpha ^{\mathrm{oh}}=0.1$. It can
be seen that, for the low-frequency bath, the anti-Zeno effect appears as
shown in Figs.~3 and 4. For the high-frequency cutoff Ohmic bath, the Zeno
effect always dominates and no anti-Zeno effect occurs. Also we can see,
from Figs. 3 and 4, that $\gamma (\tau )$ and$\ \gamma _{\mathrm{RWA}}(\tau
) $ approach $\gamma _{0}$ when the measurement interval $\tau \rightarrow
\infty $. In particular, if $\tau \rightarrow 0$, $F(\omega ,\tau
)\rightarrow 0$. Thus, $\gamma (\tau )\rightarrow 0$. This implies that in a
sufficiently short time interval of a projective measurement, the quantum
Zeno effect occurs, regardless of the bath spectrum. When the interval $\tau
$ increases, the projection interval modulation function $F(\omega ,\tau )$
displays a number of oscillations. Then, the energy peak of the bath
spectrum will act on the decay rate and it is possible to implement the
anti-Zeno effect. However, this result still depends on the function $%
f(\omega )$ and the given spectral density $J(\omega )$ of the bath. Now, we
emphasize again that the second term in the bracket of the function $%
f(\omega )=\left[ 1-(\omega -\eta \;\Delta )/\left( \omega +\eta \;\Delta
\right) \right] ^{2}$ is due to the\ counter-rotating terms; when neglecting
the counter-rotating terms, $f(\omega )=1$.

For low-frequency noise, the noise mainly comes from the region $\omega
<\Delta $, so $f(\omega )>1$. Here, $f(\omega )$ as well as $F(\omega ,\tau
) $ \textit{magnify} the effect of the energy peak of the bath spectrum.
Thus, the counter-rotating terms \textit{accelerate} the decay and the
anti-Zeno effect occurs.

In the high-frequency cutoff Ohmic bath, the noise mainly comes from the
region $\omega >\Delta $, which leads to $f(\omega )<1$. Thus, it is mainly
the counter-rotating term in $f(\omega )$ that reduces the effect of the
energy peak of the bath on the decay. This \textit{slows down the decay} and
only the quantum Zeno effect can now take place. The
projection-intervals-modulating function $F(\omega ,\tau )$, together with
the interaction-modulating function $f(\omega )$, causes the Zeno effect to
dominate in the high-frequency cutoff Ohmic bath.

\begin{figure}[tbp]
\includegraphics[width=9cm,clip]{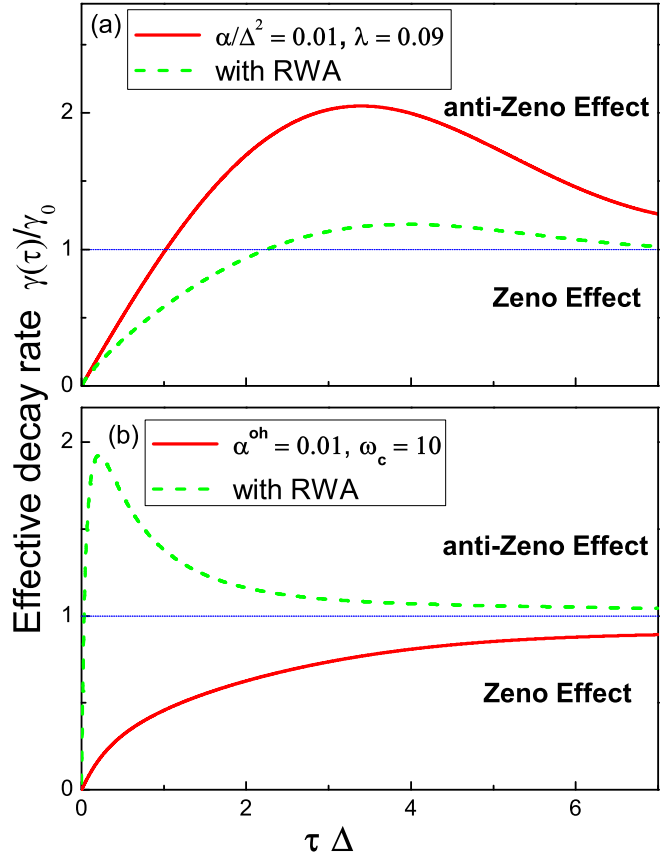}
\caption{(Color online) The effective decay $\protect\gamma (\protect\tau )/%
\protect\gamma _{0}$, versus the time interval $\protect\tau$ between
consecutive measurements, for a \textit{weak} coupling between the qubit and
the bath. In the horizontal axis, the time-interval $\protect\tau$ is
multiplied by the qubit energy difference $\Delta$. The curves in (a)
correspond to the case of a low-frequency bath with parameters $\protect%
\alpha /\Delta ^{2}=0.01$ and $\protect\lambda =0.09\Delta$ (red solid
curve). (b) corresponds to the case of an Ohmic bath with parameters $%
\protect\alpha^{\mathrm{oh}}=0.01$ and $\protect\omega _{c}=10\Delta$ (red
solid curve). The green dashed-dotted curves are the results under the RWA
when the same parameters are used. Note how different the RWA result is in
(b), especially for any short measurement interval $\protect\tau$.}
\end{figure}

\begin{figure}[tbp]
\includegraphics[width=9cm,clip]{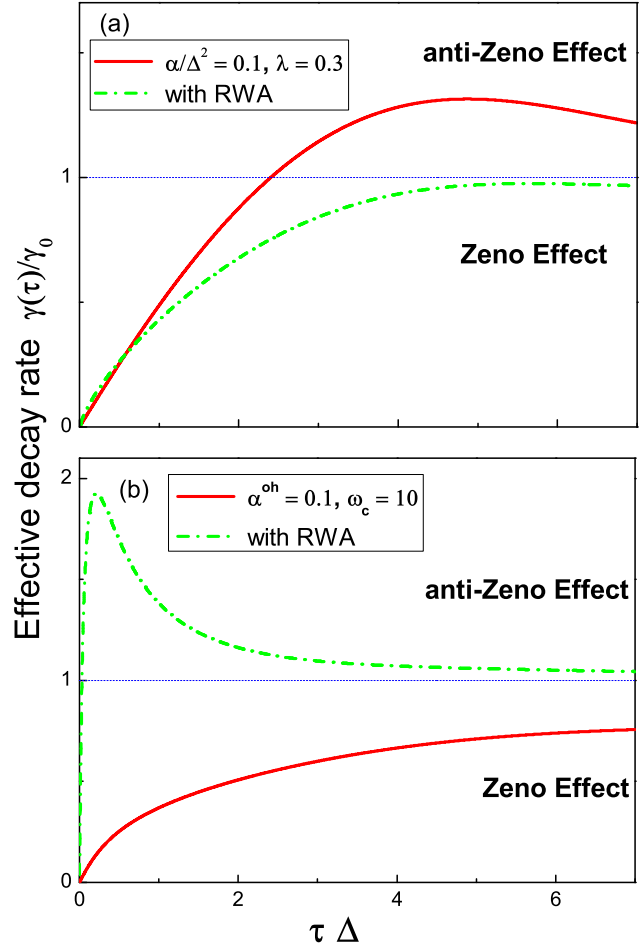}
\caption{(Color online) The effective decay $\protect\gamma (\protect\tau )/%
\protect\gamma _{0}$, versus the time interval $\protect\tau$ between
successive measurements, for a \textit{strong} coupling between the qubit
and the bath. The time-interval $\protect\tau$ is multiplied by the qubit
energy difference $\Delta$. The curves in (a) correspond to the case of a
low-frequency bath with parameters $\protect\alpha /\Delta ^{2}=0.1 $ and $%
\protect\lambda =0.3\Delta$ (red solid curve). (b) corresponds to the case
of an Ohmic bath with parameters $\protect\alpha^{\mathrm{oh}}=0.1$ and $%
\protect\omega _{c}=10\Delta$ (red solid curve). The green dashed-dotted
curves are the results under RWA when the same parameters are used. Note how
different the RWA result is in (b), especially for any short measurement
interval $\protect\tau$.}
\end{figure}

\section{Summary}

In summary, we have studied a model of a qubit interacting with its
environment, modeled either as a low- or as a high-frequency bath. For each
type of bath, the quantum dynamics of the qubit without measurement and the
quantum Zeno effect on it are shown for the cases of weak and strong
couplings between the qubit and the environment. Our results show that, for
a low-frequency bath, the qubit energy increases (blue shift) and the
quantum anti-Zeno effect occurs. However, for a high-frequency cutoff Ohmic
bath, the qubit energy decreases (red shift) and the quantum Zeno dominates.
Moreover, for a high-frequency environment, the decay rate should be faster
(without measurements) or slower (with frequent measurements, in the Zeno
regime), compared to the low-frequency bath case. These very different
behaviors of the quantum dynamics and the Zeno effect in different baths
should be helpful to experimentally distinguish the type of noise affecting
the qubit and protect the coherence of the qubit through modulating the
dominant frequency of its environment.

\vskip 0.5cm

{\noindent {\large \textbf{Acknowledgements}}}

FN acknowledges partial support from the National Security Agency (NSA),
Laboratory for Physical Sciences (LPS), Army Research Office (USARO),
National Science Foundation (NSF) under Grant No. 0726909, and JSPS-RFBR
under Contract No. 06-02-91200. X.-F. Cao acknowledges support from the
National Natural Science Foundation of China under Grant No. 10904126 and
Fujian Province Natural Science Foundation under Grant No. 2009J05014. J.-Q.
You acknowledges partial support from the National Natural Science
Foundation of China under Grant No. 10625416, the National Basic Research
Program of China under Grant No. 2009CB929300 and the ISTCP under Grant No.
2008DFA01930.

\appendix

\section*{Appendix: Examining the validity of the unitary transformation}

\label{sec:appendix}

\setcounter{equation}{0} \renewcommand{\theequation}{A.\arabic{equation}}

In this appendix, we show the main results of the canonical transformation
to the qubit-bath Hamiltonian $H$ considered here, and prove that the
contribution of $H_{2}^{^{\prime }}$ to physical quantities is of the order $%
\mathcal{O}$$\left( g_{k}^{4}\right) $ and higher, so we ignore $%
H_{2}^{^{\prime }}$ in the calculations. We now apply a canonical
transformation to the Hamiltonian $H$:
\begin{equation}
H^{\prime }=\exp \left[ S\right] H\exp \left[ -S\right] ,
\end{equation}%
with
\begin{equation}
S=\sum_{k}\frac{g_{k}}{2\omega _{k}}\xi _{k}(a_{k}^{+}-a_{k})\sigma _{x}.
\end{equation}%
Here a $k$-dependent variable, $\xi _{k}=\omega _{k}/(\omega _{k}+\eta
\;\Delta ),$ is introduced in the transformation.
It is clear that this canonical transformation is unitary, because $\left(
\exp \left[ S\right] \right) ^{+}=\exp \left[ -S\right] $. The transformed
Hamiltonian $H^{\prime }$ can now be decomposed in three parts:
\begin{equation}
H^{\prime }=H_{0}^{\prime }+H_{1}^{\prime }+H_{2}^{\prime },
\end{equation}%
with
\begin{equation}
H_{0}^{\prime }=-\frac{1}{2}\eta \;\Delta \sigma _{z}+\sum_{k}\omega
_{k}a_{k}^{+}a_{k}-\sum_{k}\frac{g_{k}^{2}}{4\omega _{k}}\xi _{k}(2-\xi
_{k}),  \label{E6}
\end{equation}%
\begin{equation}
H_{1}^{\prime }=\sum_{k}\eta \;\Delta \frac{g_{k}\xi _{k}}{\omega _{k}}%
\left( a_{k}^{+}\sigma _{-}+a_{k}\sigma _{+}\right) ,  \label{E7}
\end{equation}%
\begin{eqnarray}
H_{2}^{\prime } &=&-\frac{1}{2}\Delta \sigma _{z}\left\{ \cosh \left[
\sum_{k}\frac{g_{k}}{\omega _{k}}\xi _{k}\left( a_{k}^{+}-a_{k}\right) %
\right] -\eta \right\}  \notag \\
&&-i\frac{\Delta }{2}\sigma _{y}\left\{ \sinh \left[ \sum_{k}\frac{g_{k}}{%
\omega _{k}}\xi _{k}\left( a_{k}^{+}-a_{k}\right) \right] -\eta \sum_{k}%
\frac{g_{k}}{\omega _{k}}\xi _{k}\left( a_{k}^{+}-a_{k}\right) \right\} ,
\label{E8}
\end{eqnarray}%
where%
\begin{equation}
\eta =\exp \left[ -\sum_{k}\frac{g_{k}^{2}}{2\omega _{k}^{2}}\xi _{k}{}^{2}%
\right] .  \label{E9}
\end{equation}%
\ \ \ Note that no approximation was used during the transformation, so $%
H^{\prime }=\exp \left[ S\right] H\exp \left[ -S\right] $ is exact. Because
the constant term $\sum_{k}\frac{g_{k}^{2}}{4\omega _{k}}\xi _{k}(2-\xi
_{k}) $ in Eq.~(\ref{E6}) has no effect on the dynamical evolution, we
neglect it. Considering at low temperatures, the multiple-step process is so
weak that all the higher-order terms can be neglected. In the following
derivation, we will prove that the contribution of $H_{2}^{^{\prime }}$ to
physical quantities is of the order $\mathcal{O}$$\left( g_{k}^{4}\right) $
and higher, so we ignore $H_{2}^{^{\prime }}$ and obtain the effective
transformed Hamiltonian $H^{\prime }=H_{0}^{\prime }+H_{1}^{\prime }$.

Now let us expand the first term of $H_{2}^{^{\prime }},$ $\cosh \left[
\sum_{k}\frac{g_{k}}{\omega _{k}}\xi _{k}(a_{k}^{+}-a_{k})\right] $, as a
series in $g_{k}\xi _{k}/\omega _{k}.$ We define $\chi _{k}$ as $\chi
_{k}=g_{k}\xi _{k}/\omega _{k},$ which is proportional to $g_{k}$:
\begin{eqnarray}
\cosh \left[ \sum_{k}\frac{g_{k}}{\omega _{k}}\xi _{k}(a_{k}^{+}-a_{k})%
\right] &=&\cosh \left[ \sum_{k}\chi _{k}(a_{k}^{+}-a_{k})\right]  \label{A1}
\\
&=&\frac{1}{2}\left\{ \exp \left[ \sum_{k}\chi _{k}(a_{k}^{+}-a_{k})\right]
+\exp \left[ -\sum_{k}\chi _{k}(a_{k}^{+}-a_{k})\right] \right\} .
\label{A2}
\end{eqnarray}%
The first term in Eq.~(\ref{A2}) can be written as
\begin{equation}
\exp \left[ \sum_{k}\chi _{k}(a_{k}^{+}-a_{k})\right] =\exp \left[
\sum_{k}\chi _{k}a_{k}^{+}\right] \exp \left[ \sum_{k}-\chi _{k}a_{k}\right]
\exp \left[ -\frac{1}{2}\left[ \sum_{k}\chi _{k}a_{k}^{+},\sum_{k}-\chi
_{k}a_{k}\right] _{-}\right] ,  \label{A3}
\end{equation}%
where $\left[ \sum_{k}\chi _{k}a_{k}^{+},\sum_{k}-\chi _{k}a_{k}\right] _{-}$
means the commutator of two operators. Using the commutation relation $\left[
a_{k}^{+},a_{k}\right] =-1$, it is simplified to
\begin{equation}
\exp \left[ -\frac{1}{2}\left[ \sum_{k}\chi _{k}a_{k}^{+},\sum_{k}-\chi
_{k}a_{k}\right] _{-}\right] =\exp \left[ -\frac{1}{2}\sum_{k}\chi _{k}^{2}%
\right] =\exp \left[ -\sum_{k}\frac{g_{k}^{2}}{2\omega _{k}^{2}}\xi
_{k}{}^{2}\right] =\eta ,  \label{A4}
\end{equation}%
which is the definition of $\eta $.

Afterwards, we expand Eq.~(\ref{A3}) as follows,%
\begin{eqnarray}
\exp \left[ \sum_{k}\chi _{k}(a_{k}^{+}-a_{k})\right] &=&\eta \exp \left[
\sum_{k}\chi _{k}a_{k}^{+}\right] \exp \left[ \sum_{k}-\chi _{k}a_{k}\right]
\label{A5} \\
&=&\eta \left[ 1+\sum_{k}\chi _{k}a_{k}^{+}+\frac{\left( \sum_{k}\chi
_{k}a_{k}^{+}\right) ^{2}}{2}+\frac{\left( \sum_{k}\chi _{k}a_{k}^{+}\right)
^{3}}{3!}+...\right]  \label{A6} \\
&&.\left[ 1-\sum_{k}\chi _{k}a_{k}+\frac{\left( -\sum_{k}\chi
_{k}a_{k}\right) ^{2}}{2}-\frac{\left( \sum_{k}\chi _{k}a_{k}\right) ^{3}}{3!%
}+...\right]  \label{A7} \\
&=&\eta \left[ 1+\sum_{k}\chi _{k}a_{k}^{+}-\sum_{k}\chi
_{k}a_{k}-\sum_{k}\chi _{k}a_{k}^{+}\sum_{k}\chi _{k}a_{k}+...\right]
\label{A8}
\end{eqnarray}%
Now we see the second-order terms in $\chi _{k}$ in Eq.~(\ref{A8}):
\begin{equation}
\sum_{k}\chi _{k}a_{k}^{+}\sum_{k}\chi _{k}a_{k}=\sum_{k}\chi
_{k}^{2}a_{k}^{+}a_{k}(\mathrm{diagonal})+\sum_{k\neq k^{^{\prime }}}\chi
_{k}a_{k}^{+}\chi _{k^{\prime }}a_{k^{\prime }}(\text{off-diagonal}).
\label{A9}
\end{equation}%
The off-diagonal terms are related to the multi-boson transition and their
contributions to the physical quantities are fourth order in $\chi _{k}.$
Furthermore, the initial state $\left\vert 0_{k}\right\rangle \left\vert
\uparrow \right\rangle $ is used in the calculation, so the direct effect of
the second-order diagonal term on physical quantities is zero and its effect
through the interaction will also be fourth order in $\chi _{k}$. Thus we
now ignore the terms higher than second order in $\chi _{k}$ in the
following calculation and obtain the expression of Eq.~(\ref{A3}),
\begin{equation}
\exp \left[ \sum_{k}\chi _{k}(a_{k}^{+}-a_{k})\right] \approx \eta \left[
1+\sum_{k}\chi _{k}a_{k}^{+}-\sum_{k}\chi _{k}a_{k}\right] .
\end{equation}%
Similarly, the second term in Eq.~(\ref{A2}) becomes%
\begin{eqnarray}
\exp \left[ -\sum_{k}\chi _{k}(a_{k}^{+}-a_{k})\right] &=&\eta \exp \left[
-\sum_{k}\chi _{k}a_{k}^{+}\right] \exp \left[ \sum_{k}\chi _{k}a_{k}\right]
\\
&\approx &\eta \left[ 1-\sum_{k}\chi _{k}a_{k}^{+}+\sum_{k}\chi _{k}a_{k}%
\right] .
\end{eqnarray}%
Therefore, Eq.~(\ref{A2}) is reduced to
\begin{eqnarray}
\cosh \left[ \sum_{k}\frac{g_{k}}{\omega _{k}}\xi _{k}(a_{k}^{+}-a_{k})%
\right] &=&\cosh \left[ \sum_{k}\chi _{k}(a_{k}^{+}-a_{k})\right] \\
&=&\frac{1}{2}\left\{ \exp \left[ \sum_{k}\chi _{k}(a_{k}^{+}-a_{k})\right]
+\exp \left[ -\sum_{k}\chi _{k}(a_{k}^{+}-a_{k})\right] \right\} \\
&\approx &\eta .
\end{eqnarray}%
In the same way, the third term in $H_{2}^{^{\prime }}$ is simplified to%
\begin{equation}
\sinh \left[ \sum_{k}\frac{g_{k}}{\omega _{k}}\xi _{k}(a_{k}^{+}-a_{k})%
\right] =\eta \left[ \sum_{k}\frac{g_{k}}{\omega _{k}}\xi
_{k}(a_{k}^{+}-a_{k})+O(\chi _{k}^{3})\right] .
\end{equation}

In $H_{2}^{\prime },$ we have subtracted the terms of zero and first-order
in $g_{k},$ which are included in $H_{0}^{\prime }+H_{1}^{\prime },$ and
only left the terms equal to and higher than second-order in $g_{k},$ whose
contribution to the physical quantities is of order $\mathcal{O}$$\left(
g_{k}^{4}\right) $ and higher. Thus, $H_{2}^{\prime }$ can be omitted.

We expanded several series in the variable $\chi _{k},$ with the real
variable $\eta \Delta \chi _{k}$ less than $g_{k}$. That is: $\eta \Delta
\chi _{k}=\eta \Delta g_{k}\xi _{k}/\omega _{k}=$ $\eta \Delta g_{k}/\left(
\omega _{k}+\eta \Delta \right) <g_{k}.$ In other words, through this
transformation we find a variable smaller than $g_{k}$ for the series
expansion. Therefore, our method can be extended to the case of strong
interaction between the qubit and the environment.

\bigskip

--------------------

\end{document}